\documentclass[12pt]{article}
\usepackage{amsmath}
\usepackage{amssymb,amsthm}
\usepackage{color}

\title{Equilibrium and non-equilibrium properties of superfluids and superconductors}
\author{Walter F. Wreszinski\\
        Instituto de Fisica USP\\
        Rua do Mat\~{a}o, s.n., Travessa R 187\\
        05508-090 S\~{a}o Paulo, Brazil\\
        \texttt{wreszins@gmail.com}
        }
\begin{document}
\maketitle
\begin{abstract}
We review some rigorous results on the equilibrium and non-equilibrium properties of superfluids and superconductors. 
\end{abstract}

\section{\textbf{Introduction and motivation}}

The exact or rigorous study of equilibrium and non-equilibrium properties of superconductors has been initiated by G. L. Sewell in the 1980's. We refer to chapter 7, part III, of his book \cite{Sewell} for a beautiful introduction, and references to his own work on the subject, in particular \cite{Sewel} for the first proof of the Meissner effect in the bulk under the assumption of off-diagonal long-range order (ODLRO). In \cite{SeW} the implications of ODLRO in superfluidity were explored, including spontaneous symmetry breaking (SSB). For a discussion of the relationship between ODLRO and SSB, see the very comprehensive book by Verbeure \cite{Ver}, and for a recent study of the connection of those properties with Bogoliubov's method of quasiaverages \cite{BoBo}, see \cite{WZa2}.

Both theories, of superfluidity and superconductivity, owe greatly to the work of Bogoliubov (see \cite{BoBo} for a pedagogic exposition). The seminal work on superfluidity \cite{NNB} proposed a new approximation scheme for the weakly interacting Bose gas. For a review, with several new results, of the mathematical results of this theory, see \cite{ZBru}. A discussion of the non-equilibrium properties of superfluids also appear in chapter 3 of Bogoliubov's masterful lectures on quantum statistics \cite{Bogo}, see also \cite{Galas}. The approach to superconductivity appears in chapter 2 of \cite{Bogo}, as well as in \cite{BoTolShi}, to which we refer for references: this work had great influence on Haag's fundamental paper \cite{HaSup}, in particular concerning the theta vacua, see also \cite{WZa2} and section 5.

A recent, very stimulating historical review by Kadanoff \cite{Kad} highlights the coherence properties of the (''slippery'') wave-functions in both the phenomena of superfluidity and superconductivity, a subject well-treated in textbooks from the phenomenological point of view (see \cite{MaRo}, chapter 5.2 - the London theory). The microscopic basis of the London theory \cite{LoLo} is the property of ODLRO \cite{Sewell}. We shall come back to this point briefly when dealing with superconductivity and the Meissner effect, but otherwise shall not discuss coherence properties any further.

Superfluidity is the property of certain fluids, typically liquid Helium II, of flowing along pipes of extremely small diameter ($10^{-2}$ cm or smaller) with no resistence (viscosity). Kadanoff, in a recent historical review \cite{Kad} sharply questioned the relevance of Landau's criterion to the superfluid property. The latter may be roughly stated in the following way: by the flow of a fluid along a pipe, momentum may be lost to the walls only if the modulus of the velocity $|\vec{v}|$ is greater than
$$
v_{c} \equiv {\rm min}_{\vec{p}}\frac{\epsilon(\vec{p})}{|\vec{p}|} \, , 
$$
where $\epsilon(\vec{p})$ are the energies of the ``elementary excitations'' generated by friction. This indicates that the concept of superfluidity still lacks a clear and precise theoretical foundation. One reason, argued by Kadanoff, is that ''given the many mechanisms for broadening the distributions of both energy and momentum, it seems very implausible that the Landau condition can begin to account for the very long-lived nature of the flow of superfluid Helium''.

(Low temperature) superconductivity is the property of certain materials (e.g. mercury, tin, lead, aluminium) of becoming perfectly conducting. This is characterized by \textbf{persistent currents}, i.e., ideally everlasting currents in superconducting rings, in the absence of external magnetic fields. Together with the phenomenon of persistent currents, the prototypical electrodynamic property of superconductors, now in the presence of an external magnetic field,, is the \textbf{Meissner effect} (\cite{MaRo}, Chapter 5.2), one of the most spectacular effects in physics. It may be stated as follows:

\textbf{Meissner effect} If a superconductor sample is submitted to a magnetic (applied) field $\vec{H}$ and then cooled to a temperature $T$ below the transition temperature, there is a critical field $H_{c}(T)>0$ such that, if $|\vec{H}| < H_{c}(T)$, the field is expelled from all points of the sample situated sufficiently far from from the surface, i.e., beyond a certain distance from the surface called the ''penetration depth''.

Both (low temperature) superconductivity and superfluidity occur at very low temperatures, typically a few degrees Kelvin. 

Much less understood than the coherence properties, and still a conceptual challenge in several respects is the issue of whether the  properties of superfluids and superconductors mentioned above, are equilibrium or non-equilibrium properties \cite{Wre1}. This issue is also relevant to a third class of phenomena, that of \textbf{magnetism}: all three have in common the fact that they are macroscopic (i.e, due to infinite number of degrees of freedom) manifestations of quantum mechanics. Magnetism (diamagnetism) is also relevant to superconductivity, in the Meissner effect, and both ferro- and antiferromagnetism are relevant to certain aspects of superconductivity \cite{SewAF}. For non-equilibrium models of (disordered and ordered) ferromagnetism, see section 6.5.3 of \cite{MWB} and references given there. 

Since the thesis of Niels Bohr (1912) and the work of van Leeuwen \cite{vLee}, paramagnetism and diamagnetism were both known to require a quantum mechanical description \textbf{as equilibrium phenomena}, because the partition function has the property
$$
Z(\vec{A}) = \int d\vec{x} d\vec{p} \exp(-\beta H_{\vec{A}}(\vec{x},\vec{p})=Z(\vec{A}=\vec{0})
$$
if $H_{\vec{A}}(\vec{x},\vec{p})= \frac{(\vec{p}-e \vec{A})^{2}}{2m} + V(\vec{x})$, by a canonical transformation.

We now ask: is the flow of a fluid in superfluidity or the persistent currents in superconductivity an equilibrium or a non-equilibrium phenomenon? 

In this review we should like to explain the exact results known about this problem. Since the natural framework for this study is the theory of infinite systems, which requires the use of algebras of operators, not well-known to a large portion of the community of theoreticians, we shall dwell mostly on the conceptually familiar finite systems, with the results on algebras of operators refered to and (hopefully) briefly clarified, without requiring from the reader any previous knowledge about them, except if they wish to master the details. The main focus will lie on the concepts, and their precise definition, which is a specially subtle matter in this connection. Certain equilibrium-non-equilibrium aspects (including the Sirugue-Winnink argument), and a simpler discussion of the application to the Girardeau model, are new. 

A crucial requirement assumed on the models treated here is Galilean covariance. This is the basic symmetry of many body systems, which is, unfortunately, broken by both the Bogoliubov model and the BCS model. For a different approach, see \cite{Suto}.

\section{\textbf{General formalism, finite and infinite systems}}

We consider Bosons in translational motion (Fermions, although essential to superconductivity, will not need to be described explicitly, for reasons which will be explained later), to begin with in finite regions, which we take to be cubes, generically denoted by $\Lambda$.  The thermodynamic limit will be taken along the sequence of cubes
$$
\Lambda_{n}=[-nL,nL]^{d} \mbox{ of side } L_{n}=2nL \mbox{ with } n=1,2,3, \cdots
\eqno{(1.1)} 
$$
where $d$ is the space dimension. The number $L$ is arbitrary, and the sequence $\{n=1,2,3,\cdots \}$ may be replaced by any subset of the set of positive integers. Let 
$$
{\cal H}_{\Lambda_{j}} = L^{2}_{per}(\Lambda_{j}) \mbox{ for } j=1,2,3,\cdots
\eqno{(1.2)}
$$
denote the Hilbert space consisting of functions on $\mathbf{R}^{d}$, with $f\in L^{2}(\Lambda_{j})$ and such that f is periodic with period $L_{j}=2jL \mbox{ with } j=1,2,3,\cdots$ in each of the variables $\vec{x}=(x_{i}) \mbox{ with } i=1,\cdots,d$, i.e.,
$f(x_{i}+2jL)=f(x_{i}) \mbox{ with } i=1,\cdots,d \mbox{ and } \vec{x}=(x_{i}), i=1,\cdots,d$.
Let
$$
{\cal F}_{\Lambda_{j}} = \otimes_{s,N_{j}} {\cal H}_{\Lambda_{j}} \mbox{ with } j=1,2,3,\cdots
\eqno{(1.3)} 
$$
i.e., the symmetrized tensor product of  ${\cal H}_{\Lambda_{j}}$ corresponding to $N_{j}$ particles in $\Lambda_{j}$. 
Our local algebras of observables will be taken as the (von Neumann) algebras of bounded operators on ${\cal F}_{\Lambda_{j}}$, 
where 
$$
N_{j}/|\Lambda_{j}|=\rho
\eqno{(1.4)}
$$
with $\rho$ denoting the density and $|\Lambda|$, the volume of $\Lambda$):
$$
{\cal A}_{\Lambda_{j}} =  {\cal B}({\cal F}_{\Lambda_{j}})
\eqno{(1.5)} 
$$
Such operators may be generated by the so-called Weyl operators, see \cite{Ver}.
By the above choice one has the isotony property
$$
{\cal A}_{\Lambda_{k}} \subset {\cal A}_{\Lambda_{l}} \mbox{ for } k<l \mbox{ or } \Lambda_{k} \subset \Lambda_{l}
\eqno{(1.6)}
$$
Our states will be assumed to be functionals of the local observables which on ${\cal A}_{\Lambda}$ reduce to normal states, i.e., 
states of the form
$$
\omega_{\Lambda}(A) = {\rm Tr}_{{\cal H}_{\Lambda}}(\rho_{\Lambda} A) \qquad \forall A \in {\cal A}_{\Lambda} \, ,
\eqno{(1.7)} 
$$
where $\rho_{\Lambda}$ is a density matrix, i.e., a positive, normalized trace-class operator on ${\cal F}_{\Lambda}$; for the ground state, 
$\rho_{\Lambda} = |\Omega_{\Lambda} \rangle \langle \Omega_{\Lambda}|$, where $|\Omega_{\Lambda} \rangle$ is a normalized vector in 
${\cal F}_{\Lambda}$. In general, in the following,  ${\cal F}_{\Lambda}$ will denote the symmetrized $N$-fold tensor product of ${\cal H}_{\Lambda}$ corresponding to $N$ particles in $\Lambda$ with fixed density $\rho$. A ground state will be denoted by 
\begin{eqnarray*}
\omega_{\Lambda} = \langle \Omega_{\Lambda}, \cdot \, \Omega_{\Lambda} \rangle \, , 
\end{eqnarray*}$$\eqno{(1.8)}$$
and a temperature state 
\begin{eqnarray*}
\omega_{\beta,\Lambda} = \frac{{\rm Tr}_{{\cal F}_{\Lambda}}(\exp(-\beta H_{\Lambda})\cdot)}{Z_{\Lambda}(\beta)} \, , 
\qquad \beta >0 \, , 
\end{eqnarray*}$$\eqno{(1.9)}$$
where
\[
Z_{\Lambda}(\beta) \equiv {\rm Tr}_{{\cal F}_{\Lambda}} \exp(-\beta H_{\Lambda}) \, . 
\]
The generator of space translations on ${\cal F}_{\Lambda}$ --- the total momentum --- will be denoted by 
$\vec{P}_{\Lambda}$. Above, $\Omega_{\Lambda}$ is a ground state eigenvector of $H_{\Lambda}$ on ${\cal H}_{\Lambda}$:
\begin{eqnarray*}
H_{\Lambda} \Omega_{\Lambda} = E_{\Lambda} \Omega_{\Lambda} \, , 
\end{eqnarray*}$$\eqno{(1.10)}$$
where
\begin{eqnarray*}
E_{\Lambda} \equiv \inf {\rm spec} (H_{\Lambda}) \, . 
\end{eqnarray*}$$\eqno{(1.11)}$$
In addition,
$$
\vec{P}_{\Lambda} \Omega_{\Lambda} = \vec{0} \, .
\eqno{(1.12)}
$$
By thermodynamic stability,
$$
E_{\Lambda} \ge - c |\Lambda| \, , 
$$
where $|\Lambda|$ is the volume of $\Lambda$, and $c$ is a positive constant. In general, $E_{\Lambda}$ is of order of
$O(-d|\Lambda|)$ for some $d>0$, and in order to obtain a physical Hamiltonian satisfying positivity it is necessary to perform 
a renormalization (infinite in the thermodynamic limit)
$$
H_{\Lambda} \to \widetilde{H}_{\Lambda} \equiv H_{\Lambda} - E_{\Lambda} \, .
\eqno{(1.13)} 
$$
 Define
\begin{eqnarray*}
\alpha_{\Lambda,t}(A) \equiv \exp(itH_{\Lambda}) A \exp(-itH_{\Lambda}) \\
\sigma_{\Lambda,\vec{x}}(A) \equiv \exp(i\vec{x}\cdot\vec{P}_{\Lambda}) A \exp(-i\vec{x}\cdot\vec{P}_{\Lambda})
\end{eqnarray*}$$\eqno{(1.14)}$$
We now indicate how Galilean transformations act on finite systems.
In $\Lambda$ we consider a generic conservative system of $N$ identical particles of mass $m$. In units in which 
$\hbar = m = 1$, $H_{\Lambda}$ and $\vec{P}_{\Lambda}$ take the standard forms
$$
H_{\Lambda} = \frac{-\sum_{r=1}^{N} \Delta_{r}}{2} + V(\vec{x}_{1}, \ldots , \vec{x}_{N})
\eqno{(1.15)}
$$
with $V$ a suitable potential (satisfying certain conditions which will be specified later) and
$$
\vec{P}_{\Lambda} = -i \sum_{r=1}^{N}\nabla_{r}
\eqno{(1.16)}
$$
with usual notations for the Laplacean $\Delta_{r}$ and the gradient $\nabla_{r}$ acting on the coordinates of the r-th particle.
We assume that $H_{\Lambda}$ and $\vec{P}_{\Lambda}$ are self-adjoint operators acting on 
${\cal F}_{\Lambda}$, with domains $D(H_{\Lambda})$ and 
$D(\vec{P}_{\Lambda})$, and 
$$
D(\vec{P}_{\Lambda}) \supset D(H_{\Lambda}) \, .
\eqno{(1.17)} 
$$
Let
$$
S_{\Lambda}^{d} \equiv \left\{ \tfrac{2\pi\vec{n}}{L}\mid \vec{n} \in \mathbb{Z}^{d} \right\}
$$
and, given $\vec{v} \in \mathbb{R}^{d}$, let $\vec{v}_{\vec{n}_{L},L} = \vec{k}_{\vec{n}_{L},L}$ such that
\begin{eqnarray*}
|\vec{k}_{\vec{n}_{L},L}-\vec{v}| = \inf_{\vec{k} \in S_{\Lambda}^{d}}|\vec{k} - \vec{v}| \\
\mbox{and} |\vec{k}_{\vec{n}_{L},L}| \le |\vec{v}| \, . 
\end{eqnarray*}
If there is more than one $\vec{k}_{\vec{n}_{L},L}$ satisfying the above, we pick any one of them. We shall refer to this briefly as the \textbf{prescription}. We have:
\[
\lim_{N,L \to \infty}\vec{v}_{\vec{n}_{L},L} = \vec{v} \, , 
\]
where $N,L \to \infty$ will be always taken to mean the thermodynamic limit, whereby
\[
N \to \infty \, , \quad L \to \infty \, , \quad \frac{N}{L^{d}} = \rho \quad \mbox{ with } 0<\rho<\infty \, , 
\]
where $\rho$ is a fixed density. The unitary operator of Galilei transformations appropriate to velocity $\vec{v}_{\vec{n}_{L},L}$ follows (upon restriction to the $N$-particle subspace of symmetric Fock space):
\begin{eqnarray*}
U_{\Lambda}^{\vec{v}} \equiv \exp \left(i \vec{v}_{\vec{n}_{L},L} \cdot (\vec{x}_{1} + \ldots + \vec{x}_{N}) \right) \, 
 \end{eqnarray*}$$\eqno{(1.18)}$$
We shall assume that $U_{\Lambda}^{\vec{v}}$ maps $D(H_{\Lambda})$ into $D(H_{\Lambda})$. From now on we 
shall write $\vec{v}$ for $\vec{v}_{\vec{n}_{L},L}$.
It follows that, on $D(H_{\Lambda})$,
\begin{eqnarray*}
(U_{\Lambda}^{\vec{v}})^{\dag} \widetilde{H}_{\Lambda} U_{\Lambda}^{\vec{v}}=\\ 
H_{\Lambda,\vec{v}} + \Delta E_{\vec{v}}(\Lambda)  \, , 
\end{eqnarray*}
$$\eqno{(1.19)}$$
where
$$
H_{\Lambda,\vec{v}} \equiv \widetilde{H}_{\Lambda} +\vec{v} \cdot \vec{P}_{\Lambda}
\eqno{(1.20)}
$$
and 
$$
\Delta E_{\vec{v}}(\Lambda) \equiv \frac{N(\vec{v})^{2}}{2}
\eqno{(1.21)}
$$

We shall need to describe infinite systems and their corresponding states. This necessity arises from the fact that 1) only in this limit a simple description of the system is possible, independent of the peculiarities of finite systems (boundary conditions, etc.) 2) except in rare cases, for which surface terms become relevant even for very large systems (not met in the present article), this description is physically sound. 

For this purpose we define a so-called quasi-local agebra, which, in our case, due to the isotony property (1.6), is the limit as $k \to \infty$, of ${\cal A}_{\Lambda_{k}}$ in a suitable topology, which will be the weak topology, for reasons explained in \cite{DuSe}. We shall denote this algebra by ${\cal A}$. The (infinite-volume) states will be the thermodynamic limit of states (1.7), which will be denoted by $\omega$, with a subscript $\beta$ or none, indicating whether we are dealing with thermal states (1.9) or ground states (1.8). They have the properties of positivity $\omega(A^{*}A) \ge 0$ and normalization $\omega(\mathbf{1})=1$, inherited from (1.8) and (1.9), and are taken as defining properties of a state of an infinite system. An automorphism of an operator algebra is a one-to-one mapping of the algebra in itself, which preserves the algebraic structure. For the finite algebras the time- and space-translation automorphisms are given by (1.14). We assume that in the thermodynamic limit they give rise to automorphisms $\alpha_{t}$ and $\sigma_{\vec{x}}$ of ${\cal A}$. 

We shall be interested in time-translation invariant states $\omega(\alpha_{t}(A))= \omega(A)$ and homogeneous or space-translation invariant states: for all $\vec{x} \in \mathbf{R}^{d}$ we have that $\omega \circ \alpha_{\vec{x}} = \omega$.

A state $\omega$ is associated to a triple $(\Omega_{\omega}, \pi_{\omega}({\cal A}), {\cal H}_{\omega})$ by the GNS construction, see \cite{Sewell1}, pg. 27, where $\pi_{\omega}$ is a representation of ${\cal A}$ on a Hilbert space ${\cal H}_{\omega}$. 

For time-translation and space-translation invariant states $\omega$, in terms of this representation, $\pi_{\omega}(\alpha_{t})(A)) = \exp(itH_{\omega}) \pi_{\omega}(A) \exp(-itH_{\omega})$ for all $A \in {\cal A}$ and similarly for the space translation automorphisms $\alpha_{\vec{x}}(A)$ $\pi_{\omega}(\alpha_{\vec{x}}(A)) = \exp(i \vec{x} \cdot \vec{P}_{\omega}) A \exp(-i \vec{x} \cdot \vec{P}_{\omega})$, where $H_{\omega}$ and $\vec{P}_{\omega}$, the generators, are the (physical) Hamiltonian and momentum, satisfying
$$
H_{\omega} \Omega_{\omega} = 0
\eqno{(1.22)}
$$
and
$$
\vec{P}_{\omega} \Omega_{\omega} = \vec{0}
\eqno{(1.23)}
$$
 
Finally, we have that $\alpha_{t} \circ \alpha_{\vec{x}} = \alpha_{\vec{x}} \circ \alpha_{t}$, that is, the space and time translation automorphisms commute. We refer to \cite{Sewell1}, chapter 1, for a specially comprehensive introduction to the algebraic description of many-body (infinite) systems.  

After the renormalization (1.13), which is only possible if the finite-volume hamiltonian is bounded below (semibounded), we have $\widetilde{H}_{\Lambda} \ge 0$, which leads in the thermodynamic limit to the condition
$$
H_{\omega} \ge 0
\eqno{(1.24)}
$$
(1.22) also implies that $\omega$ is $\alpha_{t}$- invariant. Concerning thermal states, the finite volume Gibbs states (1.9) satisfy certain analyticity properties known as the Kubo-Martin-Schwinger (KMS) boundary condition, see again \cite{Sewell1} for a comprehensive account. Although for a finite system and a given dynamics $\alpha_{\Lambda,t}$ (see (1.14)), there is only one Gibbs state (1.9), this is not so for infinite systems, for which phase transitions may occur. It is therefore important to characterize equilibrium states of infinite systems by a more general condition \cite{Sewell1}, \cite{Hug}, \cite{BRo2} :let $0 < \beta <\infty$ denote the inverse temperature. A state $\omega$ is a $(\alpha_{t},\beta)$ - KMS state if, $\forall A,B \in {\cal A}$, there exists a function $F_{A,B}$ analytic inside the strip ${\cal D}_{\beta} \equiv \{z | 0 < \Im z <\beta\}$, bounded and continuous on its closure $\overline{{\cal D}_{\beta}}$, and satisfying the KMS boundary condition
\begin{eqnarray*}
F_{A,B}(t) = \omega(A\alpha_{t}(B)) \mbox{ and } F_{A,B}(t+i\beta) = \omega(\alpha_{t}(B)A) \mbox{ for all } t\in\mathbb{R} \, . 
\end{eqnarray*}$$\eqno{(1.25)}$$
   
A KMS state is $\alpha_{t}$-invariant.

A state is a ground state if it is $\alpha_{t}$- invariant and satisfies the spectrum condition (1.24), and an equilibrium state at temperature $\beta$ if it is a $\beta$- KMS state.

Although thermodynamic limits of Gibbs states satisfy the KMS condition, it is not implied that such are the only ones. In fact, the KMS condition may itself be used to construct thermal states for infinite systems, as done by van Hemmen \cite{LvHem} for BCS and related models of mean field type.

\section{ A physical characterization of superfluidity and superconductivity} 

The essential physical characterization of superfluid and superconductive states is the fact that they carry a current \cite{Sewell}. One envisages, typically, the following idealized situation.

A superfluid or superconductor, initially in an equilibrium state (thermal or ground state), is set into motion by an external agent, e.g. a pressure gradient between two points of a pipe in the former case, a voltage gradient in the latter, which imparts to the system a steady macroscopic current, represented by a classical vector field: we refer to this situation as ''the system in motion''. The main question we ask is whether the initial equilibrium state is also an equilibrium state of the system in motion.  

For simplicity, we consider first superfluids in a fixed inertial frame, the restframe of the system in the (equilibrium) ground state $\Omega_{\Lambda}$. We assume that it is Galilean covariant, i.e., (1.19)-(1.21).  The energy which must be imparted to the system to set it into motion with uniform velocity $\vec{v}$ is the expectation value in $\Omega_{\Lambda}$ of
\begin{eqnarray*}
U_{\Lambda}^{\vec{v}})^{\dag} \widetilde{H}_{\Lambda} U_{\Lambda}^{\vec{v}} -\widetilde{H}_{\Lambda}\\
= \vec{v} \cdot \vec{P}_{\Lambda} + \Delta E_{\vec{v}}(\Lambda)
\end{eqnarray*}
which equals $\Delta E_{\vec{v}}(\Lambda)$. According to Landau, one should compare this energy with the eigenvalues (or ''elementary excitations'') of the Hamiltonian operator of the system in motion with velocity $\vec{v}$, namely, $U_{\Lambda}^{\vec{v}})^{\dag} \widetilde{H}_{\Lambda} U_{\Lambda}^{\vec{v}}$, which amounts to  analyse the spectrum of $H_{\vec{v},\Lambda}$ , given by (1.20) : if it is positive for a given range of $\vec{v}$, no dissipation occurs. For the relation of the above to the concepts of passivity and semipassivity, see the article by B. Kuckert \cite{Kuc}.

By Galilean covariance, the momentum operator for the system in motion is given by
$$
\vec{P}_{\Lambda,\vec{v}} \equiv \vec{P}_{\Lambda} + N \vec{v}
\eqno{(1.26)}
$$
By (1.12), the system carries a current 
$$
\vec{J}_{\Lambda} \equiv N \vec{v}
\eqno{(1.27.1)}
$$
which is the steady macroscopic current alluded to above. The system in motion is thus described by the Hamiltonian $H_{\Lambda,\vec{v}}$ given by (1.20). Due to (1.12), $\Omega_{\Lambda}$  is also an eigenstate of $H_{\Lambda,\vec{v}}$ and therefore the former equilibrium state is also invariant under the dynamics of the system in motion. This fact generalizes to the thermal states (1.9), because  $H_{\Lambda,\vec{v}}$ commutes with $H_{\Lambda}$, as well as to the thermodynamic limits of both ground and thermal states.

The same considerations apply to the superconductor with a current, as remarked by Feynman in his lectures (\cite{Feyn}, 10.9). Instead of (1.27.1) we have $$
\vec{J}_{\Lambda} \equiv -N e\vec{v}            
\eqno{(1.27.2)}
$$
where $-e$ denotes the charge of an electron pair, and the excess energy, i.e, the energy furnished by the external agent (voltage) is $\Delta E_{\vec{v}}(\Lambda)$ ((10.56) and (10.57) of \cite{Feyn}).

\section{\textbf{Nonequilibrium phenomena in superconductivity and superfluidity}}

We wish to inquire whether the initial equilibrium states of the system remain so in the presence of motion. For finite systems, this is not possible, both for thermal states and for ground states. This does not, however, imply the same result for infinite systems. A Gibbs state at temperature $\beta$ given by (1.9) cannot be a Gibbs state at the same temperature for the dynamics defined by (1.20), which we denote $\omega_{\beta,\Lambda,\vec{v}}$. This follows from the fact that the requirement $\omega_{\beta,\Lambda}(A) = \omega_{\beta,\Lambda,\vec{v}}(A)$ for all $A \in {\cal A}_{\Lambda}$ implies $\vec{v}=\vec{0}$ by choosing $A= \exp(-\beta \widetilde{H}_{\Lambda})-\exp(-\beta H_{\Lambda,\vec{v}})$, which is possible by our choice of algebra, which is weakly closed. From the condition that $\lim \omega_{\Lambda}(A) = \lim \omega_{\Lambda,\vec{v}}(A)$, for all $A \in {\cal A}_{\Lambda}$, where $\lim$ denotes the thermodynamic limit, the conclusion does not follow, but we shall see shortly that, formulating the equilibrium condition as the KMS condition (1.25), the conclusion does follow.

The reader may think that the above is a minor point of rigor, but perhaps the argument for ground states will convince him of the contrary. For finite systems, the ground state condition $\widetilde{H}_{\Lambda} \ge 0$ does not imply the analogous condition $H_{\Lambda,\vec{v}} \ge 0$ for the system in motion.
This is a consequence of
$$
spec(H_{\Lambda,\vec{v}}) = spec(\widetilde{H}_{\Lambda}) - N \vec{v}^{2}/2
\eqno{(1.27.3)}
$$
which follows from the unitary transformation (1.19) together with the prescription. Indeed, $H_{\Lambda,\vec{v}}$ and $\widetilde{H}_{\Lambda}$ share a common complete set of eigenvectors, independently of the prescription, as one sees from (1.20), among which $U_{\Lambda}^{-\vec{v}}|\Omega_{\Lambda}$ is the ground state of $H_{\Lambda,\vec{v}}$. Indeed, by (1.20), and (1.12),
$$
H_{\Lambda,\vec{v}} U_{\Lambda}^{-\vec{v}}|\Omega_{\Lambda}) = -1/2 N |\vec{v}|^{2}U_{\Lambda}^{-\vec{v}}|\Omega_{\Lambda})
\eqno{(1.27.4)}
$$ 
As remarked in \cite{SeW}, all  this is, however, not sufficient to show that the infinite system in motion has a physical Hamiltonian $H_{\omega} + \vec{v} \cdot \vec{P}_{\omega}$  which is not positive in the representation determined by $\omega$. (1.27.3-4) suggest that, ''in normal circumstances'', $H_{\vec{v},\Lambda}$ has a spectrum contained in $[-\Delta E_{\vec{v}}(\Lambda),\infty)$, which tends to the whole real line as $N,L \to \infty$. In order to see what may go wrong, assume that the spectrum ${\rm spec} \, (H_{\vec{v},\Lambda}) = [-\Delta E_{\vec{v}}(\Lambda), - \lambda_{N,L}]\cup[0,\infty)$, with $\lambda_{N,L} \to \infty$ as $N,L \to \infty$. In this case, the negative part of the spectrum ''disappears'' in the thermodynamic limit, and the assertion that the spectrum of the physical Hamiltonian (for the infinite system) contains a non-empty set in the negative real axis does not follow.
 
A further hint that (1.27.3-4) do not imply that the infinite system in motion has a physical Hamiltonian $H_{\omega} + \vec{v} \cdot \vec{P}_{\omega}$  which is not positive in the representation determined by $\omega$ is given by the fact that, in the thermodynamic limit, the state $ U_{\Lambda}^{-\vec{v}}|\Omega_{\Lambda})$ gives rise to a representation inequivalent (not unitarily equivalent, more precisely disjoint) to the one associated to $\omega$. This is a well-known phenomenon, intrinsic to infinite number of degrees of freedom, see chapter 1 of \cite{Sewell1} or \cite{MWB}, corollary 6.3, pg. 255: two different values of $\vec{v}$, in particular $\vec{v}$ and $\vec{0}$, correspond to two different macroscopic densities, viz. momentum densities, by (1.26).

It is therefore necessary to show that the spectrum of $H_{\Lambda, \vec{v}}$ covers the whole negative real axis in the thermodynamic limit, a fact which, as we shall see in the case of superfluidity, has a deep physical origin.

In order to proceed, we consider both dynamics, one related to the inertial system (1.14) and the other to the system in motion. For a finite system, the latter's  corresponding automorphisms are given by
\begin{align*}
\alpha_{t,\vec{v},\Lambda}(A) & = {\rm e}^{itH_{\vec{v}}(\Lambda)} A {\rm e}^{-itH_{\vec{v}}(\Lambda)} \qquad
\forall A \in {\cal A}_{\Lambda} \, , 
\end{align*}$$\eqno{(1.28)}$$
and their infinite volume version will be denoted by $\alpha_{t,\vec{v}}$. For space translations, by (1.26), the automorphisms $\sigma_{\vec{x}}$ are the same.

\subsection{\textbf{A general approach to non-equilibrium states: the Sirugue-Winnink argument}}

We assume that $\omega$ is time- and space-translation invariant, i.e.,
$$
\omega(\alpha_{t}(A)) = \omega(A)
\eqno{(1.29)}
$$
and
$$
\omega(\sigma_{\vec{x}}(A)) = \omega(A)
\eqno{(1.30)}
$$
where $\alpha_{t}$ and $\sigma_{\vec{x}}$ are commuting automorphisms of ${\cal A}$.
Let $\omega$ denote a state of a Galilean covariant Boson or Fermion system, $\omega =(\Omega, \cdot \, \Omega)$ and
$$
H_{\omega} \Omega = 0 \mbox{ and }  \vec{P}_{\omega} \Omega = \vec{0} 
\eqno{(1.31)}
$$
Define the self-adjoint operators
$$
H_{\vec{v},\omega} \equiv H_{\omega} + \vec{v} \cdot \vec{P}_{\omega}
\eqno{(1.32)} 
$$
for $\vec{v} \in \mathbb{R}^{d}$. The corresponding automorphism group of ${\cal A}$ is the $\alpha_{t,\vec{v}}$ referred to above. We provide a sketch of proof of the following theorem, for those who only wish to understand the main argument.

\textbf{Theorem 1}
Let $ \omega =\omega_{\beta}$ satisfy the KMS condition with respect to $\alpha_{t}$ for a given $0<\beta <\infty$, where $\alpha_{t}= \alpha_{t,\vec{0}}$. Then, the quantum dynamical system $(\omega_{\beta},\alpha_{t,\vec{v}},{\cal A})$ defines a non equilibrium stationary state (NESS) whenever $\vec{v} \ne \vec{0}$.

\textbf{Proof} We must show that $\omega_{\beta}$ does not satisfy the KMS condition  
with respect to the dynamics defined by $\alpha_{t,\vec{v}}$ .
\textbf{Sketch of proof}:
Assume the contrary, i.e., that $\omega_{\beta}$ satisfies the KMS condition with respect to 
$\alpha_{t,\vec{v}}$. Since it also satisfies the KMS condition with respect to 
$\alpha_{t}$ by hypothesis, applying the KMS condition first w.r.t.~$\alpha_{t}$ and then w.r.t.~$\alpha_{t,\vec{v}}$, 
we find, for $A \in {\cal A}$, 
\begin{align*}
F_{A,B}(t) \equiv \omega_{\beta}(\alpha_{t}(A)\alpha_{t,\vec{v}}(B)) 
& = \omega_{\beta}(\alpha_{t,\vec{v}}(B)\alpha_{t+i\beta}(A)) \nonumber \\
& =\omega_{\beta}(\alpha_{t+i\beta}(A)\alpha_{t+i\beta,\vec{v}}(B)) \nonumber \\
& =F_{A,B}(t+i\beta) \, . 
\end{align*}
$$\eqno{(1.33)}$$
(1.33) leads straightforwardly to the fact that $F_{A,B}$ is analytic in the whole complex plane and is therefore a constant
which equals $\omega_{\beta}(AB)$, the value of $F_{A,B}$ at $t=0$.
Varying $A$ and $B$ we generate the whole Hilbert space and, from the definition of $F$, it follows that $(U_{t,\omega_{\beta}})^{-1} U_{t,\omega_{\beta},\vec{v}} = \mathbf{1} \, ,$, from which 
$H_{\omega} + \vec{v} \cdot \vec{P}_{\omega} = H_{\omega}$ and, thus, $\vec{v} = \vec{0}$. 

We now provide a complete proof, for those who wish to delve on the details.

We actually obtain (1.33) only for $A \in {\cal A}_{\alpha_{t}}$, $B \in {\cal A}_{\alpha_{t,\vec{v}}}$, where ${\cal A}_{\tau}$ is a norm-dense *-subalgebra of 
${\cal A}$ consisting of entire analytic elements for $\tau$ (see \cite{BRo1}, Definition 2.5.20 and Proposition 2.5.22),
The function $F_{A,B}(z)$ is, for $A \in {\cal A}_{\alpha_{t}}$, $B \in {\cal A}_{\alpha_{t,\vec{v}}}$, analytic in 
${\cal D}_{\beta} = \{z \in \mathbb{C}\mid0<\Im z <\beta\}$ and continuous on the closure $\overline{{\cal D}}_{\beta}$. By the three-line 
lemma (see, e.g., \cite{BRo2}, Proposition 5.3.5), it is uniformly bounded in $\overline{{\cal D}}_{\beta}$ by
$\|A\|\| B\| $. Choosing sequences $\{A_{n}\}_{n \ge 1},\{B_{n}\}_{n \ge 1}$ with $A_{n} \in {\cal A}_{\alpha_{t}}$, 
$B_{n} \in {\cal A}_{\alpha_{t,\vec{v}}}$, such that
$$
\| A_{n}\|  \le \| A\| ; \quad \lim_{n \to \infty}\| A_{n}-A\|  = 0 \, , 
$$
$$
\| B_{n}\|  \le \| B\| ; \quad \lim_{n \to \infty}\| B_{n}-B\| = 0 \, , 
$$
one obtains that $F_{A_{n},B_{n}}(z) \to F_{A,B}(z)$ uniformly in $\overline{{\cal D}}_{\beta}$ (see \cite{BRo2}, pg.~82), the limit function being 
therefore continuous and bounded on $\overline{{\cal D}}_{\beta}$, analytic in~${\cal D}_{\beta}$, and satisfying the periodicity condition (1.33):
$$
F_{A,B}(t) = F_{A,B}(t+i\beta) \qquad  \forall t \in \mathbb{R} \, , \quad \forall A,B \in {\cal A} \, . 
$$
Furthermore, the following uniform bound holds as a consequence of the previous inequalities , which themselves follow from Kaplansky's density 
theorem, Theorem 2.4.16 of \cite{BRo1}:
$$
|F_{A,B}(z)| \le \| A\| \| B\|  \qquad \forall z \in {\cal D}_{\beta} \, . 
\eqno{(1.34)}
$$
Hence, by the Schwarz reflection principle, the function $F_{A,B}$ extends uniquely to an analytic function in the whole of $\mathbb{C}$, 
satisfying the bound (1.34), and is therefore a constant. From the definition (1.33), it follows that 
$F_{A,B}(z) = \omega_{\beta}(AB)$ for all $z \in \mathbb{C}$, whence the unitaries $U_{t,\omega_{\beta}}$ and $U_{t,\omega_{\beta},\vec{v}}$
implementing the automorphisms 
$\alpha_{t}$ and $\alpha_{t,\vec{v}}$
in the GNS representation of $\omega_{\beta}$, satisfy
$$
(U_{t,\omega_{\beta}})^{-1} U_{t,\omega_{\beta},\vec{v}} = \mathbf{1} \, , 
$$
upon using the cyclicity of $\Omega_{\beta}$. By our choices of $\alpha_{t}$,$\alpha_{t,\vec{v}}$, this implies that 
$\vec{v}= \vec{0}$. q.e.d.

\textbf{Remark} The above proof simplifies the original theorem in \cite{SiWi}.

The above approach  identifies what seems to be a central element in the issue equilibrium vs non-equilibrium in the present context: the \textbf{same} state and \textbf{two different} dynamics under which the state is assumed invariant (note that KMS states are necessarily invariant if the dynamics are implemented by an automorphism). The latter point accounts for the stationarity of the non-equilibrium states.

\subsection{\textbf{The role of time reversal invariance: Bloch's theorem}}

We now assume that the equilibrium state is time-reversal invariant, i.e., there exists an antiautomorphism $\theta$ of the observable algebra such that $\omega \circ \theta = \omega$. Considering the representation $\pi_{\omega}$ associated to $\omega$ by the GNS construction, assume that $\vec{j}(f)$ is a self-adjoint operator on a domain in ${\cal H}_{\omega}$ which includes $\Omega_{\omega}$, and which has the same property as the currents appearing in many-body quantum mechanics, i.e. $U_{\theta} \vec{j}_(f)U_{\theta}^{-1} = -\vec{j}(f)$ where the antiunitary operator $U_{\theta}$ implements $\theta$ in the GNS representation. Then we have

\textbf{Bloch's theorem} $\omega(\vec{j}(f)) = \vec{0}$

Indeed, $\omega(\vec{j}(f)) = (\omega \circ \theta)(\vec{j}(f))= \omega((\theta(\vec{j}(f))))= -\omega(\vec{j}(f)) = \vec{0}$. 

Theorem 1 generalizes Bloch's theorem because it is an assertion on the state, not only restricted to the expectation value of the current. It follows from both theorems that the persistent currents in superconductivity are definitely a \textbf{non-equilibrium} phenomenon.

\subsection{\textbf{Nonequilibrium nature of translational superfluids: the ground state}}

Theorem 1 is valid for strictly positive temperatures, but not for the ground state. Moreover, the general basic assumption that the time and space translations define automorphisms of the (weakly closed) quasilocal algebra, and even the existence of the physical Hamiltonian $H_{\omega}$ and physical momentum $\vec{P}_{\omega}$ , very natural for any system, have been proved only for the free Bose gas and the BCS model \cite{DuSe}. It is therefore better \cite{Wre1} to base one's discussion on the Green's function for the finite system \cite{BRo2}, defined as
\begin{eqnarray*}
G_{\Lambda,\vec{v}}(A,B;t,\vec{x})\equiv \omega_{\Lambda}(A \alpha_{\Lambda,\vec{v},t}(\sigma_{\Lambda,\vec{x}}(B))\\
\forall A,B \in {\cal A}_{L} \mbox{ and } t \in \mathbf{R}
\end{eqnarray*}

It may be shown \cite{BRo2},\cite{Wre1} that there exist subsequences $\Lambda_{n_{k}}$ such that the limits 
\begin{eqnarray*}
G_{\vec{v}}(A,B;t,\vec{x}) = \lim_{k} G_{\Lambda_{n_{k}},\vec{v}}(A,B;t,\vec{x}) \\
\forall A,B \in {\cal A} \forall t \in \mathbf{R} \forall \vec{x} \in \mathbf{R}^{d}
\end{eqnarray*}
exist. Thus, a sequence $\Lambda_{n_{k}}$ such that the thermodynamic limit of the Green's functions  
exists can always be found but is not unique: such nonuniqueness is related to the existence of phase transitions, which are expected for Boson systems at very low temperatures.
 
We are now interested in considering the state $\omega$, which is an equilibrium state at zero temperature (ground state), under the different dynamics, whose generator is $H_{\vec{v},\omega}$.  It is time-translation invariant, under this dynamics, but it is not necessarily an equilibrium state; precisely this situation has been studied before for KMS states. 

We have that $|\Omega_{\Lambda})$  is an eigenstate of $H_{\vec{v},\Lambda}$ of eigenvalue zero, so that stationarity is guaranteed and, indeed, follows naturally from the physical input without the need of performing a time-averaging on the state $\omega$.

We now define the \emph{energy-momentum spectrum} $emsp_{N_{k},L_{k}}$ of the finite system as 
the set of pairs $(E_{r,N_{k},L_{k}}(\vec{v}), \vec{k}_{s,L_{k}})_{r,s}$, where $\{E_{r,N_{k},L_{k}}(\vec{v})\}_{r}$ denotes the set of eigenvalues of $H_{\vec{v},\Lambda_{k}}= H_{\vec{v},N_{k},L_{k}}$, and 
$\{\vec{k}_{s,L_{k}}\}_{s} = S_{\Lambda_{k}}^{d}$  the set of eigenvalues of the momentum $\vec{P}_{N_{k},L_{k}}$. Above, when $\vec{v}$ occurs in the context of the finite system, the prescription is always understood.

\textbf{Definition 1} $emsp_{\infty}^{l}(\vec{v})$ is the set of limit points of $emsp_{N_{k},L_{k}}$, possibly along subsequences.
 
In \cite{Wre1} we defined the following concept (see, e.g., \cite{BB}, for the concept of support of a distribution):

\begin{eqnarray*}
emsp_{\infty}(\vec{v}) \equiv \mbox{ The support of the Fourier transform of } \\
G_{\vec{v}}(A^{*},A;t,\vec{x}) \mbox{ as a tempered distribution }
\end{eqnarray*}

We also showed that it agrees with the support of the joint spectral family $E(\lambda,\vec{k})$ of $(H_{\vec{v},\omega},\vec{P}_{\omega})$  in case these generators of time and space translations for the system in motion exist, and, furthermore, that
$$
emsp_{\infty}^{l}(\vec{v}) \subset emsp_{\infty}(\vec{v})
\eqno{(1.35)} 
$$
This leads to

\textbf{Definition 2} $emsp_{\infty}(\vec{v})$ will be called the energy-momentum spectrum of the infinite system. If 
$\exists (\lambda,\vec{k}) \in emsp_{\infty}(\vec{v})$ such that $\lambda < 0$, we shall say that the system describes a NESS.

The last statement of definition 2 is justified by the fact that, when generators exist, a zero-temperature state is a non-equlibrium state iff $spec(H_{\vec{v},\omega}) \cap (-\infty,0) \ne \emptyset$ by (1.24). Stationarity is included because, when generators exist, the corresponding state is time-translation invariant.

\textbf{Corollary 1}
The corresponding ground state of the infinite system is a NESS If $emsp_{\infty}^{l}(\vec{v})$ contains a point $(\lambda(\vec{k_{0}}),\vec{k_{0}})$ with $\lambda(\vec{k}_{0}) < 0$.

\subsection{\textbf{Metastability: Landau superfluids as NESS}}

Searching, now, for a metastability condition, i.e., a condition which might account for the long-livedness of the superfluid state, we are led to define subspaces of states ''close'' to the superfluid state, restricted to which $H_{\vec{v},\Lambda}$ is expected to be positive.
A natural proposal, connected to Kadanoff's remark in the introduction related to the broadness in energy and momentum, is the following. Let ${\cal H}_{N,L}^{c,d}$ denote the subspace of ${\cal H}_{N,L}$  generated by of all linear combinations of vectors $\sum_{i,j} \lambda_{i,j} \Psi_{i,j}$  with  $\Psi_{i,j}$  simultaneous eigenvectors of $\widetilde{H}_{N,L}$ and $ P_{N,L}$ corresponding to eigenvalues  $E_{i,N,L}(\vec{v}=\vec{0})$ and $\vec{k}_{j,L}$ such that 
$$
E_{i,N,L} \le c \mbox{ for all } i \mbox{ and } |\vec{k}_{j,L}| \le d
\eqno{(1.36.1)}
$$

The restrictions in  on the magnitude of the energy-momentum around the ground state above are due to the fact that we wish to express that there should be a sufficiently small energy-momentum transfer between the particles of the fluid and the surroundings (pipe). As remarked by Baym \cite{Ba}, the assumption that the energy of the excitations (in our case: the spectrum of $H_{\vec{v}}$) does not depend on the velocity of the walls is implicit. This assumption should be valid, but only for slow relative motion of the superfluid and walls.

When speaking of pipe and walls we assume that the periodic b.c. is not true in all directions, for dimensions greater than one. This means that $\Lambda$ is not a cube, and the total momentum and velocity are replaced by $\vec{P}_{\Lambda} = -i \sum_{r=1}^{N} \frac{\partial}{\partial x_{r}} \vec{e}_{1}$ and 
$\vec{v} = \pm v \vec{e}_{1}$, respectively, and $\vec{v}$ is the velocity of the wall. 

Accordingly, we pose:

\textbf{Definition 3 - A metastability condition} 
The system defines a \emph{superfluid} at $T=0$ if $\exists 0<v_{c}<\infty$ such that, whenever $|\vec{v}| \le v_{c}$, 
there exists a subspace ${\cal R}_{N,L}^{c,d}$ of 
${\cal H}_{N,L}^{c,d}$ such that the eigenvalues $\{E_{j,N,L}(\vec{v})\}_{j}$ of the restriction of
$\widetilde{H}_{N,L} + \vec{v}_{\vec{n}_{L},L} \cdot \vec{P}_{N,L} = H_{\vec{v},N,L}$ to
${\cal R}_{N,L}^{c,d}$ satisfy
$$
\epsilon^{c,d}_{j}(\vec{v}) \equiv \lim_{N,L \to \infty} E_{j,N,L}(\vec{v}) \ge 0 \mbox{ for all } j
\eqno{(1.36.2)}
$$
and that
$$
\{\epsilon^{c,d}_{j}(\vec{v})\}_{j} \ne \{0\}
\eqno{(1.36.3)}
$$
for some $0<c<\infty$, $0<d<\infty$.

The last condition is a non-triviality requirement, which is expected to be satisfied in general (and is so for the Girardeau-Lieb-Liniger model). 

Definition 3 is related to Conjecture 2.2 of Cornean, Derezinski and Zin \cite{CDZ}. It may be expected to be a general characterization of metastability for superfluid Boson systems, independently of the presence or absence of BEC. An important issue is that definition 3 goes beyond \textbf{thermodynamics}, and for this reason Cornean et al \textbf{define} the infimum of the energy spectrum (IES) in the thermodynamic limit $e^{\rho}(\vec{k})$, where $\rho$ again denotes the density. We adopted a definition somewhat closer to the algebraic spirit, but both approaches have, of course, specific advantages.

\subsection{\textbf{An illustrative example: the Girardeau-Lieb-Liniger model}}

We start with the Girardeau-Lieb-Liniger model \cite{Gir}, \cite{LLi} of $N$ particles (for simplicity odd) in one dimension with repulsive delta function 
interactions, whose formal Hamiltonian is given by (mass $m=1/2$)
$$
H_{N,L} = -\sum_{i=1}^{N}\frac{\partial^{2}}{(\partial x)^{2}} + 2c (\sum_{i,j=1}^{N})^{'} \delta(x_{i}-x_{j}) 
\mbox{ with } 0 \le x_{i},x_{j} \le L  \, . 
\eqno{(1.37)}
$$
The prime over the second sum indicates that the sum is confined to nearest neighbors. For the (standard) corresponding rigorous definition 
, see \cite{Do}.

The limit, as $ c \to \infty$, of equation (4), yields Girardeau's model \cite{Gir}. It is the free particle Hamiltonian with Dirichlet b.c.~on 
the lines $x_{i}=x_{j}$, with quadratic form domain given in (\cite{Do}, pg.~353, (2.18)-(2.19)). It is straightforward that $U_{\Lambda}^{\vec{v}}$ leaves this domain invariant and (1.17,1.19) hold. It is essential that we speak of the finite system, for which $H_{\vec{v},N,L}$ is bounded from below, and quadratic form methods apply; note that (1.17) is strict, because $\vec{P}_{\Lambda}$ is unbounded from below also for the finite system.
 
For $c \to \infty$, the b.c. on the wave-functions reduces to
$$
\Psi(x_{1},\ldots,x_{N}) = 0 \mbox{ if } x_{j}=x_{i}, \; 1 \le x_{i},x_{j} \le N \, ,
\eqno{(1.38.1)} 
$$
and the (Bose) eigenfunctions $\Psi^{B}$ satisfying equation (1.38.1) simplify to
$$
\Psi^{B}(x_{1},\ldots,x_{N}) = \Psi^{F}(x_{1},\ldots,x_{N})A(x_{1},\ldots,x_{N}) \, ,
\eqno{(1.38.2)} 
$$
where $\Psi^{F}$ is the Fermi wave-function for the free system of $N$ particles confined to the region 
$0 \le x_{i} < L,i=1,\ldots,N$, with periodic b.c., and
$$
A(x_{1},\ldots,x_{N}) = \prod_{j<l} sign(x_{j}-x_{l}) \, . 
$$
Note that $\Psi^{F}$ automatically satisfies equation (1.38.1) by the exclusion principle. Indicating the Bose and Fermi ground states by the 
subscript $0$, it follows from equation (1.38.2) and the non-negativity of $\Psi_{0}^{B}$ that
$$
\Psi_{0}^{B} = |\Psi_{0}^{F}| \, . 
$$
Since $A^{2} = 1$, the correspondence between $\Psi^{B}$ and $\Psi^{F}$ given by equation (1.38.2) preserves all scalar 
products, and therefore the energy spectrum of the Bose system is the same as that of the free Fermi gas. Furthermore, $\Psi^{F}$ is a 
Slater determinant of plane-wave functions labelled by wave vectors $k_{i},i=1,\ldots,N$, equally spaced over the range $[-k_{F},k_{F}]$,
where $k_{F}$ is the Fermi momentum. Hence $k_{F}=\pi\frac{N-1}{L}$, which, in the thermodynamic limit reduces to 
$$
k_{F} = \pi \rho \, .
\eqno{(1.39)}
$$
The simplest excitation is obtained by moving a particle from $k_{F}$ to $q > k_{F}$ (or from $-k_{F}$ to $q < -k_{F}$) thereby leaving 
a hole at $k_{F}$ (or $-k_{F}$). This excitation has momentum $k=q-k_{F}$ (or $-(q-k_{F})$), and energy
$\epsilon(k)=(q^{2}-k_{F}^{2})/2$, i.e.,
$$
\epsilon^{1}(k)=k^{2}/2 + k_{F}|k| \, . 
\eqno{(1.40)}
$$
This type of excitation must be supplemented by the umklapp excitations, which we consider in a more general form than \cite{Lieb}. 
They consist in taking a particle from $(-k_{F}-p)$ to $(k_{F}+q)$ or $(k_{F}-q)$ to $(-k_{F}-p)$. We call the corresponding eigenvalues 
$\epsilon^{3}(k)$ and $\epsilon^{2}(k)$ and consider just the latter in detail; for these
$$
0 \le q \le 2\pi(N-1) \mbox{ and } \frac{2\pi}{L} \le p
\eqno{(1.41.1)}
$$
their  momentum equals
$$
k=-2k_{F}-(p-q)
\eqno{(1.41.2)}
$$
and their energies $\epsilon^{2}(k)$ are given by
$$
2\epsilon^{2}(k) = [(-k_{F}-p)^{2}-(k_{F}-q)^{2}]=[2k_{F}+(p-q)](p+q) \, .
\eqno{(1.41.3)} 
$$
A different, equivalent choice involving holes may be made \cite{Lieb}.

We consider now a sufficiently large number of umklapp excitations; their momentum is given by (1.41.2) and thus opposite to $\vec{v}$ by our (arbitrary) choice (otherwise we would have taken the other set of umklapp excitations). Consider, now
$$
H_{v,N,L} = H_{N,L} + v P_{N,L}
$$
We now switch one particle from $k_{F}$ to $-k_{F}-p$ (corresponding to $q=0$, $p=\frac{2\pi}{L}$ in (1.41.1)). This corresponding elementary excitation aquires thereby the momentum 
$$
k_{1} = -2k_{F} - \frac{2\pi}{L}
\eqno{(1.41.4)}
$$
Taking now a particle from $k_{F}-\frac{2\pi}{L}$ to $-k_{F}-2\frac{2\pi}{L}$ (corresponding to $q=\frac{2\pi}{L}$, $p=2\frac{2\pi}{L}$ in (1.41.1), this again leads to the same elementary excitation momentum (1.41.4), and the resulting momentum of the two excitations is $-2(2k_{F})+o(L)$.
      
Performing now $r$ successive umklapp excitations we obtain that the corresponding  excitation energy $e(v,L)$ of $H_{v,N,L} = H_{N,L} + v P_{N,L}$ is, by (1.41.4), with $\rho=N/L$,
\begin{eqnarray*}
e(v,L)=(2k_{F}+2\pi/L)\sum_{i=1}^{r} (\frac{2\pi i}{L} - \pi/L) + v \sum_{i=1}^{r}(-2k_{F}-2\pi/L)=\\
= (2k_{F}+2\pi/L)[2\pi/L (\frac{(1+r)r}{2}) -(\pi/L) r] + v(-2k_{F}r-2\pi/L)
\end{eqnarray*}$$\eqno{(1.41.5)}$$
which is $-2k_{F}vr +o(L)$, with $k_{F}$ given by (1.39). It is specially interesting that, for the \textbf{finite} system, (1.41.5) leads to the value
$$
e(v,L)= -\frac{Nv^{2}}{4}+o(L)
$$
(notice that the mass $m=1/2$), upon choosing $r=[Lv/2\pi]$, that is, a sequence of umklapp excitations lead to the bottom of the spectrum (see (1.27.3)), up to small corrections.

By further reasoning based on (1.40) \cite{Wre1}, we obtain:

\textbf{Proposition 1} For the Girardeau model, if
$$
0< |\vec{v}|< \pi\rho 
\eqno{(1.42.1)} 
$$
the spectrum ${\rm spec} \, (H_{v,N,L})$ of $H_{v,N,L}$ contains the set ${\cal E}$, defined by
$$
{\cal E} = \{-2k_{F}vj + o(L)\} \mbox{ with } j=1,2,3, \cdots
\eqno{(1.42.2)}
$$
Thus, the quantities
$$
\lambda_{j} \equiv -2k_{F}v j \mbox{ with } j=1,2, \ldots
\eqno{(1.42.3)}
$$
belong to $emsp_{\infty}^{l}$  in the sense of definition 1, which is ,thus, unbounded from below, and therefore the system describes a NESS by corollary 1. Furthermore, adopting
$$
c= (\pi \rho)^{2}
\eqno{(1.43.1)} 
$$
and
$$
d=\pi \rho 
\eqno{(1.43.2)}
$$
in definition 3, it follows that the system describes a superfluid at $T=0$ in the sense described there.

We now explain why, with the choices (1.43) in definition 3, (1.36.2) holds on a subspace ${\cal R}_{N,L}^{c,d}$. Let  ${\cal R}_{N,L}^{c,d}$ denote the subspace of ${\cal H}_{N,L}^{c,d}$ generated by a fixed number $r$ independent of $N,L$ of elementary excitations - for simplicity of the type considered in the proof of proposition 1- as well as fixed numbers $s_{1},s_{2} \cdots$ of the remaining excitations, with the additional constraints: the momentum operator $\vec{P}_{N,L}$ has in this subspace a fixed value $\vec{P}$ such that
$$
|\vec{P}| < d=\pi \rho 
\eqno{(1.44.1)}
$$
and for the energy we require
$$
H_{N,L} - E_{N,L}^{0} \mbox{ when restricted to } {\cal R}_{N,L}^{c,d} \ge c
\eqno{(1.44.2)}
$$
with $c$ given by (1.43.1). If
$$
|\vec{k}| < \pi \rho
\eqno{(1.44.3)}
$$
where $\vec{k} = \vec{k}_{1} + \cdots + \vec{k}_{r}$, the momentum $\vec{l}$ of the remaining excitations must satisfy $|\vec{l}| = |\vec{P}-\vec{k}|<\pi \rho + \pi \rho = 2 \pi \rho$, and thus the umklapp excitations are absent from ${\cal R}_{N,L}^{c,d}$. Since all the elementary excitations have positive energies, (1.44.2) and (1.40) imply that
$$ 
k_{F} |\vec{k}| \le c
\eqno{(1.44.4)}
$$
with $\vec{k} = \vec{k}_{1}+ \cdots + \vec{k}_{r}$, for $N$ sufficiently large and fixed $r$, due to the $N^{-1}$-corrections, hence by (1.43.1) and (1.44.4), it follows that
$$
|\vec{k}| \le |\vec{k}_{1}|+ \cdots + |\vec{k}_{r}| \le \frac{c}{\pi \rho} = \pi \rho
$$
proving (1.44.3). Since, for $\vec{v}$ satisfying (1.42.1), $\epsilon^{1}(\vec{k}_{i}) + \vec{v} \cdot \vec{k}_{i} -\vec{k}_{i}^{2}/2 \ge 0$ for all $i=1, \cdots, r$, and similarly for the other excitations (excluding the umklapp excitations, which we have shown to be absent), it follows that the restriction of $H_{N,L} - E_{N,L}^{0} + \vec{v} \cdot \vec{P_{N,L}}$ to ${\cal R}_{N,L}^{c,d}$ is positive, which is (1.36.2). The fact that the subspace ${\cal R}_{N,L}^{c,d}$ does not shrink to the empty set in the thermodynamic limit, i.e., (1.36.3), is easily seen to be true for any $r \ge 1$, see \cite{Wre1}.

For $c$ sufficiently large proposition 1 holds for the Lieb-Liniger model by continuity, but it is not straightforward to prove it for arbitrary $c$, although the result is expected, because  a detailed analysis of the equations in the appendix of \cite{LLi}, not done there for the umklapp excitations, is required: there are, however, no problems of principle.

The velocity of sound $c_{s}$ derived from the excitation spectrum $c_{s} = \lim_{k \to 0} \frac{\partial e(k)}{\partial k}$ in one dimension is related to the compressibility by $c_{s} = [(\frac{-L}{m \rho}\frac{\partial P}{\partial L}]^{-1/2}$ where $P$ is the pressure $P=\frac{-\partial E_{0}}{\partial L}$. For the Girardeau model (and also for the Lieb-Liniger model, but there the argument is less simple) infinite point repulsion makes $E_{0}$ proportional to
$\sum_{p=1}^{(N-1)/2} (\frac{2 \pi p}{L})^{2}$ which is proportional to $N (N+1)(2N +1)/6 L^{-2}$. differentiating twice with respect to $L$ we get
$-L\frac{\partial p}{\partial L} = O(\frac{N^{3}}{L^{3}})$ which is nonzero in the thermodynamical limit. Comparing with free bosons, $E_{0}= N/L^{2}$ and the same calculation yields zero in the thermodynamic limit, or infinite compressibility. For this reason the free Bose gas is a bad model even in the limit of strong dilution.

\section{\textbf{Equilibrium: the Meissner effect in superconductivity}}

As described in the introduction, the Meissner effect relates to superconductors in the presence of an external magnetic field. The Hamiltonian is therefore not time-reversal invariant, and thus  Bloch's theorem is not applicable to the Meissner effect. In fact, strong arguments have been given \cite{Schaf1} that the latter is an \textbf{equilibrium phenomenon}. Of this remarkable review, \cite{Schaf1}, which appeared shortly before his untimely death in a plane crash, we also quote: ''The most serious failing of the theory of BCS is, however, its failure to account for the electrodynamical properties of superconductors, such as the Meissner effect and the persistent currents.'' The reason for this is the BCS model's lack of local gauge covariance, see \cite{Sewell}, chapter 7, the introduction to \cite{Wre2} and the footnote in \cite{BCS}. 

The model we shall revisit was studied by Schafroth \cite{Schaf3} as one of the very few locally gauge invariant systems possibly related to superconductivity. Today, it may be viewed as a model for Schafroth pairs, which are known to occur in certain compounds , i.e., quasi-bound electron pairs localized in physical space, i.e., for which the spatial extension of the pair wave function, measured by the coherence length, is small compared with the average distance between pairs. In this case all electrons of the band are paired, and the pairs form a dilute Bose gas. The present account from \cite{Wre2} omits several technical details, e.g., regularity assumptions, and concentrates on just the essential points. The Hamiltonian may be written ($e=2e_{0}$, $e_{0}$ being the electron charge):
\begin{eqnarray*}
H(\vec{A}) = \frac{\hbar^{2}}{2m} \int_{K} (\nabla + \frac{ie}{\hbar} \vec{A}(\vec{x}))a^{*}(\vec{x}) \cdot\\
\cdot (\nabla - \frac{ie}{\hbar} \vec{A}(\vec{x})) a(\vec{x}) d\vec{x}
\end{eqnarray*}$$\eqno{(1.45)}$$
where $a(\vec{x})$ and $a^{*}(\vec{x})$ are the basic destruction and creation operators on symmetric Fock space ${\cal F}_{s}({\cal H})$, with 
${\cal H} = L^{2}(K)$ the (one-particle) Hilbert space of square integrable wave functions on the cyçlinder $K$, assumed to be of radius $R$ and height $L$, centered at the origin, of volume $V= \pi R^{2} L$.

$H(\vec{A})$ is the (self-adjoint) second quantization of a one-particle operator on ${\cal H}$, with certain boundary conditions. We shall, however, use an extended version of this one-particle operator. The current density operator in this model is 
$$
\vec{j}(\vec{x}) = \vec{j}_{mom}(\vec{x}) + \vec{j}_{Lon,\vec{B}}(\vec{x})
\eqno{(1.46.1)}
$$
where
$$
\vec{j}_{mom}(\vec{x}) = -\frac{ie\hbar}{2m}:a^{*}(\vec{x})\nabla a(\vec{x}) - a(\vec{x}) \nabla a(\vec{x}):
\eqno{(1.46.2)}
$$
and
$$
\vec{j}_{Lon,\vec{B}}(\vec{x}) = -\frac{e^{2}}{m} a^{*}(\vec{x}) a(\vec{x}) \vec{A}(\vec{x})
\eqno{(1.46.3)}
$$
Above, :: denotes the Wick or normal product. $\vec{j}_{Lon,\vec{B}}(\vec{x})$ denotes the \textbf{London} part of the current.

Let $(\vec{e}_{\rho},\vec{e}_{\theta},\vec{e}_{z})$ be a positively-oriented orthonormal basis adapted to cylindrical coordinates. We assume that the magnetic induction $\vec{B}$ is uniform outside the cylinder, i.e., that the system is embedded in $\mathbf{R}^{2} \times [-L/2,L/2]$, and
$$
\vec{B}(\rho,\theta,z) = B \vec{e}_{z} \mbox{ for } \rho \ge R
$$
where $B$ is a positive constant.  The \textbf{semiclassical} (static) model will be \textbf{defined} by the above equation, together with the set of Maxwell equations for the magnetic induction $\vec{B}$ (using the MKS system):
$$
(\nabla \cdot \vec{B})(\vec{x}) = 0
$$
$$
(\nabla \times \vec{B})(\vec{x}) = \mu_{0} (\Omega_{\vec{B}}, \vec{j}(\vec{x}) \Omega_{\vec{B}})
$$
where
$$
\vec{B}(\vec{x}) = (\nabla \times \vec{A})(\vec{x})
$$
and $\Omega_{\vec{B}}$ is the ground state of $H(\vec{A})$, assumed unique (see later). Note that the g.s. $\Omega_{\vec{B}}$ depénds itself on the solution to the Maxwell equation.

At present no existence theorem for the semiclassical model is known, and we shall have to assume:

\textbf{Assumption 1}
For any $0 < B < \infty$, the semiclassical model has a solution of the form
$$
\vec{A}(\vec{x}) = \left\{
\begin{array}{rl}
\frac{B \rho}{2} \vec{e}_{\theta} & \text{if~} \rho \ge R \,\\
a(\rho) \vec{e}_{\theta}          & \text{if~} \rho \le R \,,
\end{array}
\right.
$$

The above (with $K$ now replaced by the extended region) implies that the state $|\Omega_{\vec{B}})$ is of the form
$$
|\Omega_{\vec{B}}) = \frac{\Psi^{*}(\phi_{0})^{N} |\Omega_{0})}{(N!)^{1/2}}
$$
where $|\Omega_{0})$ is the Fock vacuum (no-particle state), and $\phi_{0}$ denotes the normalized ground-state wave function of the one-particle operator
$$
H_{\vec{A}}^{1} \equiv \frac{(\vec{p} - e\vec{A}(\vec{x}))^{2}}{2m}
$$
which, in correspondence to the embedding in $\mathbf{R}^{2} \times [-L/2,L/2]$ above, will be considered as an operator on
$$
{\cal H}_{e} \equiv L^{2}(\mathbf{R}^{2},d\mu(\rho,\theta)) \otimes L^{2}((-L/2,L/2))
$$
with
$$
d\mu(\rho,\theta) = d\mu(\rho) d\theta \mbox{ with } d\mu(\rho) = \rho d\rho
$$

\textbf{Local gauge covariance} is defined by
$$
U_{\alpha}^{-1} H(\vec{A}) U_{\alpha} = H(\vec{A} + \nabla \alpha)
$$
which is readily seen  to be satisfied by the Hamiltonian $H(\vec{A})$.
The Maxwell equation is more precisely stated in the form $(\nabla \times \vec{B})(f) = \mu_{0} (\Omega_{\vec{B}}, \vec{j}(f) \Omega_{\vec{B}})$; taking a sequence $\{f_{n}^{\vec{x}}\}_{n \ge 1}$, with $f_{n}^{\vec{x}} \to \delta(\vec{x})$ in the distributional sense, we may obtain the Maxwell equation in the form
$$
(\nabla \times \vec{B})(\vec{x}) = \mu_{0} \lim_{n \to \infty} (\Omega_{\vec{B}}, \vec{j}(f_{n}^{\vec{x}}) \Omega_{\vec{B}})
$$
in case the limit on the r.h.s. above exists, which is seen to hold. The standard choice of operators $a$, $a^{*}$, implies a choice of phase $\alpha(\vec{x})=0$ or a real ground state wave function $\phi_{0}$. Choosing a different $\alpha=\alpha(\vec{x})$ instead, one obtains an additional term in the momentum density part of the current. For this reason, we should write, more precisely:
$$\vec{j}_{Lon,\vec{B}}(\vec{x}) = -\frac{e^{2}}{m} a^{*}(\vec{x}) a(\vec{x}) (\vec{A}(\vec{x})-(\nabla \alpha)(\vec{x}))$$
where $\nabla \alpha$ is a gauge function which balances the gauge non-invariance of $\vec{A}$, in order that a gauge invariant combination results.

We assume that Neumann boundary conditions are imposed at $z= \pm L/2$. $H_{\vec{A}}^{1}$ may be written
$$
H_{\vec{A}}^{1} = \bigoplus_{k \in \mathbf{Z}} H_{\vec{A}}^{1}(k)
\eqno{(1.47.1)}
$$
where
\begin{eqnarray*}
H_{\vec{A}}^{1}(k) = \frac{\hbar^{2}}{2m}(-\frac{\partial^{2}}{\partial \rho^{2}} - \frac{\partial}{\rho \partial \rho}+\\
+ \frac{|k+ \rho \alpha(\rho)|^{2}}{\rho^{2}} - \frac{\partial^{2}}{\partial z^{2}})
\end{eqnarray*}$$\eqno{(1.47.2)}$$
with $\alpha(\rho)$ given by
$$
\alpha(\rho) = \left\{
\begin{array}{rl}
\frac{e\rho B}{2\hbar}  & \text{if~} \rho \ge R \,,\\
\frac{e a(\rho)}{\hbar} & \text{if~} \rho \le R \,,
\end{array}
\right.
\eqno{(1.47.3)}
$$

The above formula for the Hamiltonian leads us to expect that for $B$ sufficiently small the ground state corresponds to $k=0$, i.e., $\phi_{0}$ is independent of $\theta$, and that only the London part of the current contributes to the r.h.s. of the Maxwell equation.

This is very difficult to prove, because $\alpha(\rho)$ also depends on this same wave function by the Maxwell equation. 

We thus pose:

\textbf{Assumption 2} There exists a constant $B_{0}$ such that, for $B< B_{0}$, $H_{\vec{A}}^{1}$, given by (1.47), has a ground state in the sector $k=0$. 

Under this assumption the eigenvalue problem reduces to that of a radial Schroedinger equation with potential $\alpha(\rho)^{2}$ and Maxwell's equation becomes an integrodifferential for $B(\rho)$, whose analysis by the contraction mapping principle in \cite{Wre2} shows that it decays exponentially inside the cylinder, with a ''penetration depth'' independent of the size of the sample and consistent with the physical data. 

This model is, therefore, not ''exactly soluble'', as originally thought to be by Schafroth \cite{Schaf3}, on account of the fact that there is a ''potential'' $\alpha(\rho)^{2}$, and, moreover, this potential is basically unknown! For this reason a proof of ODLRO for this model is very difficult.

\section{Conclusion}

We showed that current-carrying states cannot be thermal equilibrium states (theorem 1). In the case of superfluidity, it was demonstrated, together with results of \cite{Wre1}, that ground states cannot be equilibrium states if a certain condition on the limit points (possibly along subsequences) of the energy-momentum spectrum of finite systems is met (corollary 1), and that in the latter case the systems describe non-equlibrium stationary states (NESS). In the Girardeau-Lieb-Liniger model it was shown that the conditions of corollary 1 are indeed met, and that the energy spectrum of the infinite system is unbounded from below (proposition 1). In the specific case of the Girardeau model, this unboundedness from below was related to the existence of certain elementary excitations - the umklapp excitations - which derive from the existence of an effective ''Fermi distribution'' for the Boson system. The latter is, on the other hand, a direct consequence of the \textbf{repulsive} nature of the interactions, which, for realistic systems, is present in the van der Waals forces between neutral atoms. Although the  unboundedness from below in the thermodynamic limit of $H_{\Lambda,\vec{v}}$ is not expected to be specific to repulsive interactions, we know of no other rigorous examples even in the repulsive case.  
 
The umklapp excitations, which imparting large momentum and small energy to the particles, may be special to singular repulsive interactions in one-dimensional systems. Nevertheless, Seiringer and Yin \cite{SeYin} showed that the Lieb-Liniger model is a suitable limit of dilute Bosons in three dimensions, and, thus, a qualitatively similar picture may hold for more realistic systems. In fact, Lieb's two branches of excitations have been seen experimentally, see \cite{SKODTD}. In general, for realistic systems, it is conjectured that the instability for translational superfluids is caused by rotons, which arise as local minima of the infimum of the excitation spectrum, see \cite{MaRo}, pg.~246. These should also take place only for sufficiently large momentum, 
as a consequence of the properties of the liquid structure factor, if one assumes the Feynman variational wave function, see \cite{MaRo}, pg.~256. 
For the Feynman variational wave function, see \cite{CDZ} and \cite{WSi}. 

Another issue related to elementary excitations is that of metastability. The concept of local stability (of thermal states) introduced by Haag, Kastler and Trych-Pohlmeyer \cite{HKTP} is not suitable to describe stability of superfluids and superconductors. This becomes specially clear in the deep analysis of this concept in the case of ground states due to Bratelli, Kishimoto and Robinson \cite{BKR}, who showed that the condition of local stability in \cite{HKTP} when applied to the ground state is equivalent not to the ground state condition (1.24) (which would be the analogue of the equivalence condition proved - under certain assumptions - in \cite{HKTP} in the thermal case, namely, the KMS condition (1.25)), but rather to the existence of a \textbf{gap} in the spectrum of the physical Hamiltonian $H_{\omega}$. This is due to the possibility that the local perturbation creates an infinite number of \textbf{infraparticles} of infinitesimally small energy, a possibility which indeed occurs in superfluidity and superconductivity as a consequence of Goldstone Bosons arising from spontaneous symmetry breaking \cite{Ver}, \cite{SeW}, \cite{WZa2}. At the same time, this suggests a clue to a possible definition of metastability, our definition 3, i.e., stability when restricted to a subspace spanned by ''a few'' elementary excitations, essentially as suggested by Kadanoff \cite{Kad}, see also \cite{CDZ}. This is shown for the Girardeau-Lieb-Liniger model in proposition 1. In this connection, note that The ground state of $H_{\vec{v},\Lambda}$ corresponds, by (1.27.3) , to the reversion of the velocities of a macroscopic number of particles. This explains the connection to metastability: very large momentum is necessary to connect the original ground state to this state.

Lenard showed \cite{Len} that for the Girardeau model no BEC takes place. Therefore, superfluidity and BEC are conceptually different issues. However, they may be related in the presence of Bose-Einstein condensation (BEC). The analysis of an interacting model with nonzero density in three dimensions exhibiting BEC and superfluidity remains as the most important, perennial open problem.

For superconductors, there are no persistent currents in thermal equilibrium states by theorem 1, a generalization of  Bloch's theorem, which uses time-reversal invariance. Nevertheless, some of the best textbooks e.g. \cite{Zi} insist (pp.339-341) that persistent currents arise as a consequence of the gap in BCS theory (which is an equilibrium theory). The fact that Feynman's Ansatz (1.27.2) requires Galilean covariance, which is broken in the BCS theory due to the truncation done to arrive at the model, also confirms that the BCS model is not adequate to analyse this issue.
 
In spite of its lack of local gauge invariance, with serious consequences for the electrodynamics, analysed at length by Schafroth \cite{Schaf1} (see also \cite{Sewell} and \cite{Wre2}), the BCS model remains an important model in mathematical physics, and the study of its mathematical structure, starting with the seminal work of R. Haag \cite{HaSup}, has led to very important developments, in the papers of Thirring and Wehrl \cite{ThirrW} and Thirring \cite{ThirrSup}, culminating in the comprehensive and more general approach of van Hemmen \cite{LvHem}, who refers to Haag's method as the ''Bogoliubov-Haag approach'' (see the introduction). An important recent review \cite{HaSei} summarizes extensive recent progress in the study of the mathematical properties of the BCS functional.

Besides the open problem of persistent currents, which is a non-equilibrium effect, there are other effects which depend on the electrodynamics in superconductivity and which are equilibrium effects, and, among these, the Meissner effect plays a major role. Although Schafroth argues that the BCS Ansatz \cite{BCS} in their approach to the Meissner effect is actually equivalent to the assumption of the existence of the effect itself, there is no general agreement on this issue, and, Kadanoff argued in his review \cite{Kad} that the problem was solved by Anderson \cite{An5}, who proposed that oscillations of the gap parameter produces extra states of the system (plasmons) which rescue the local gauge invariance of the theory. Although this may be true physically,  it seems non-controversial that Anderson's solution is not susceptible to rigorous or exact bounds, and, in the sense of the present paper, the Meissner effect has really to be considered as an open problem, the more so that the BCS model in its original form is an exactly soluble mean-field model (see also \cite{LvHem} and references given there). 

The model of Schafroth pairs introduced by Schafroth \cite{Schaf3} to study the Meissner effect exactly, revisited in \cite{Wre2} and reviewed in this paper seems to be the simplest, yet nontrivial enough to merit further study. We have already discussed why a proof of ODLRO for this models appears as (very) difficult.  Assumptions 1 and 2 represent, therefore, the central open problem in this model: the existence of a unique solution of the London type, for sufficiently small values of the imposed magnetic field. 

\textbf{Acknowledgement} The present review was written for the meeting on Operator Algebras and Quantum Physics, July 17th-23rd 2015, Satellite conference at USP, S$\tilde{a}$o Paulo, Brazil, to the XVIII International Conference on Mathematical Physics, Santiago de Chile (2015). We should like to thank the organizers for the invitation, A. S\"{u}t\"{o} for helpful correspondence, B. Nachtergaele, G. L. Sewell and V. A. Zagrebnov for helpful remarks, and both referees, specially referee 2, for very enlightening comments.

\end{document}